\documentclass[12pt]{article}
\usepackage{epsf}

\begin{document}

\title{Chandler wobble: two more large phase jumps revealed}
\author{Zinovy Malkin and Natalia Miller \\ Central Astronomical Observatory at Pulkovo of RAS,\\ Pulkovskoe Ch. 65, St.~Petersburg 196140, Russia}
\date{23 August 2009}
\maketitle

\begin{abstract}
Investigations of the anomalies in the Earth rotation, in
particular, the polar motion components, play an important role in
our understanding of the processes that drive changes in the
Earth's surface, interior, atmosphere, and ocean.
This paper is primarily aimed at investigation of the Chandler
wobble (CW) at the whole available 163-year interval to search for
the major CW amplitude and phase variations.
First, the CW signal was extracted from the IERS (International
Earth Rotation and Reference Systems Service) Pole coordinates
time series using two digital filters: the singular spectrum analysis
and Fourier transform. The CW amplitude and phase variations were
examined by means of the wavelet transform and Hilbert transform.
Results of our analysis have shown that, besides the well-known CW phase jump in
the 1920s, two other large phase jumps have been found in the 1850s and 2000s.
As in the 1920s, these phase jumps occurred contemporarily with a sharp
decrease in the CW amplitude.
\end{abstract}

%%%%%%%%%%%%%%%%%%%%%%%%%%%%%%%%%%%%%%%%%%%%%%%%%%%%%%%%%%%%%%%%%%%%%%%%%%%%%%

\section{Introduction}
\label{intro}

The Chandler wobble (CW) is one of the main components of motion of the Earth's
rotation axis relative to the Earth's surface, also called Polar motion (PM).
It was discovered by Seth Carlo Chandler in 1891 (Chandler 1891a, 1891b). The
CW is one of the main eigenmodes of the Earth rotation, and investigation of
its properties such as period, amplitude and phase variations is very important
for the understanding of the physical processes in the Earth, including its
surface, interior, atmosphere and ocean. Since the discovery of CW, numerous
papers were devoted to analysis of this phenomenon. The results are summarized
in Munk \& MacDonald (1960) and Lambeck (1980). For the latest result see e.g.
Schuh et al. (2001) and referenced papers.

Among other interesting CW peculiarities, a phase jump of about $180^\circ$
occurred in the 1920s. Perhaps, it was for the first time detected by Orlov
(1944). Detailed consideration of this CW phase jump is given e.g. in Guinot
(1972) and Vondr\'ak (1988). Also, less significant CW phase jumps can be
observed, which may be in temporary coincidence with geomagnetic jerks and Free
Core Nutation (FCN) phase perturbations (Gibert \& Le Mou\"el 2008, cf. Shirai
et al. 2005). However, the ``main'' phase jump in the 1920s remains to be most
interesting feature of the CW, which, in particular, is a real test of Earth
rotation theories.

It should be noted that most papers devoted to the CW studies are based chiefly
on the observations of the Earth rotation obtained in the period from 1899
(start of the International Latitude Service) till the end of 1990s. Recently
fulfilled study of CW based on analysis of the whole IERS (International Earth
Rotation and Reference Systems Service) C01 PM time series (Miller 2008) has
shown clear evidence of two other large CW phase jumps that occurred in the
beginning and the end of the C01 series. Earlier, Sekiguchi (1975) found a CW
phase perturbation of about 40$^\circ$ in the 1850s, but he did not come to a
definite conclusion due to the poor quality of the observations used.

In this paper, we perform detailed investigation of the IERS Pole motion data
to finally confirm that preliminary finding. Our study consists of three steps:
forming the longest available IERS PM time series joining C01 and C04 series,
extracting CW signal, and analyzing CW signal thus obtained. To improve
reliability of the results, several methods of analysis were used, and their
results corroborate the common conclusion on the two existing large CW phase
jumps in the 1850s and nowadays.

%%%%%%%%%%%%%%%%%%%%%%%%%%%%%%%%%%%%%%%%%%%%%%%%%%%%%%%%%%%%%%%%%%%%%%%%%%%%%%

\section{Preliminary data processing}
\label{sect:data}

For our analysis, we used the EOP(IERS)C01 PM series that spans the period from
1846.0 through 2008.5. We extended this series up to 2009.0, merging it with
the EOP(IERS)C04 series. These series are sampled at different rates: 0.1 yr
(C01, 1846--1899), 0.05 yr (C01, 1890--2008), 1 d (C04). Using this data, three
163-year evenly spaced series with 0.05 yr, 0.1 yr, and 10 d step were
constructed, and all the computations described below were made with all three
series.  No visible differences in results were found.

Our next goal was to extract the CW signal from the PM time series, removing
all the regular (periodic and quasi-periodic) and non-regular (trend) PM
components out of the CW frequency band. Using digital filtering enables us to
cut off the PM variations with periods beyond the CW band without application
of any models of annual, trend, and other signals. The following two methods
were used for digital filtering of the PM series.
\begin{description}
\item[{\it Singular spectrum analysis (SSA).}] This method allows us to investigate the time series structure
   in more detail than other digital filters.  As shown in previous studies, it can be effectively used
   in investigations of the Earth rotation (see, e.g., Vorotkov et al. 2002, Miller 2008).
\item[{\it Foruier filtering.}]
   We used the bandpass Fourier transform (FT) filter with the the window 1.19$\pm$0.1 cpy.
   Such a wide filter band was used to preserve the complicated CW structure.
   In the filtered PM series, the amplitude of the remaining annual signal is about 0.5 mas, i.e. 0.5\% of the original value.
\end{description}

\begin{figure}[!ht]
\centering
\epsfclipon \epsfxsize=0.85\hsize \epsffile{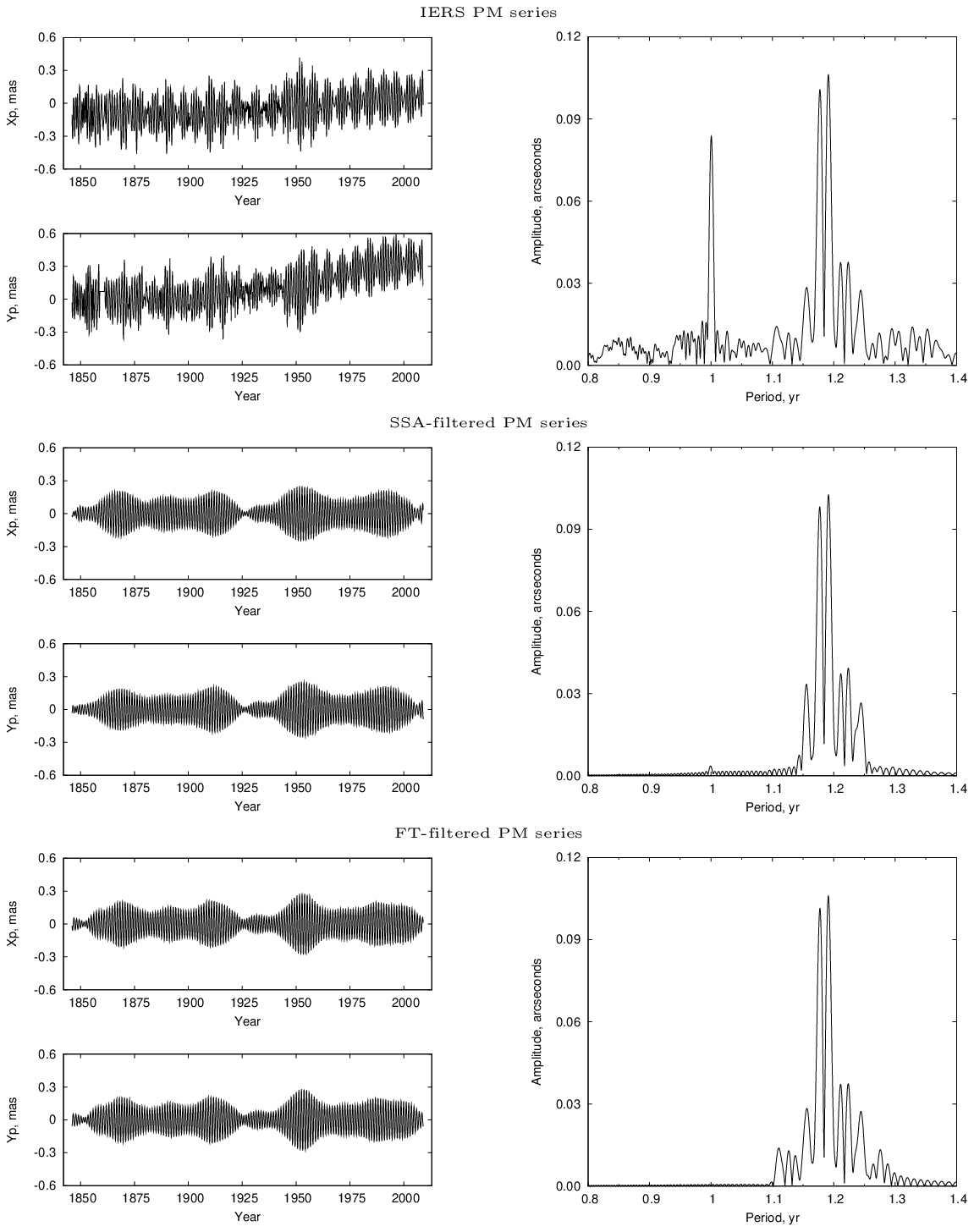}
\caption{Original and filtered PM series used for our analysis, and corresponding spectra. One can see that both types of digital
filtering allows us to effectively suppress the annual signal. The CW signal looks similar in both filtered series.
However, some differences can be seen near the ends of the interval.}
\label{fig:series}
\end{figure}

Hereafter we will refer to filtered PM time series as CW series. Analyzed PM
and CW time series and their spectra are shown in Fig.~\ref{fig:series}. We can
see two main spectral peaks of about equal amplitude near the central period of
about 1.19 yr, and several less intensive peaks in the CW frequency band.
Discussion on its origin, and even reality, lies out of the scope of this
paper. Here we can only mention that observed bifurcation of the main CW peak
is most probably caused by the phase jump in 1920s as suggested in to Fedorov
\& Yatskiv (1965). Detailed analysis of this problem is given e.g. by Sekiguchi
(1976) and Guo et al. (2005). According to our study, comparison of the spectral
components obtained at different time intervals gave us the single main CW
period $P_0=1.185$ yr.

%%%%%%%%%%%%%%%%%%%%%%%%%%%%%%%%%%%%%%%%%%%%%%%%%%%%%%%%%%%%%%%%%%%%%%%%%%%%%%

\section{Evaluation of the CW parameters}
\label{sect:parameters}

The next step of our study was to estimate the CW parameters, namely, instant
amplitude and phase. Let us consider a general CW model with variable amplitude $A(t)$
and phase $\Phi(t)$ which can be written as
\begin{equation}
\begin{array}{rcl}
X_p(t) & = & A(t)\cos\Phi(t) \,,\\
Y_p(t) & = & A(t)\sin\Phi(t) \,,
\end{array}
\label{eq:model}
\end{equation}
where $X_p$ and $Y_p$ are the Pole coordinates. Mathematically (not
geophysically, indeed!), we can suppose three equivalent models for the CW phase:
\begin{equation}
\Phi(t) = \left\{
\begin{array}{l}
\displaystyle\frac{2\pi}{P(t)}\,t + \varphi_0 \,, \\[1.5em]
\displaystyle\frac{2\pi}{P_0}\,t + \varphi(t) \,, \\[1.5em]
\displaystyle\frac{2\pi}{P(t)}\,t + \varphi(t) \,,
\end{array}
\right.
\label{eq:phase}
\end{equation}
where P is the CW period, and zero subscripts mean  constant (time-independent) values.
In other words, we can consider the following three models: with variable period
and constant phase, variable phase and constant period, or variable both period and phase.
Of course, this is a subject of a special geophysical consideration.

The CW amplitude time series can be easily computed as
\begin{equation}
A(t) = \sqrt{X_p(t)^2+Y_p(t)^2} \,,
\label{eq:amplitude}
\end{equation}

\begin{figure}[!ht]
\centering
\epsfclipon \epsfxsize=0.45\hsize \epsffile{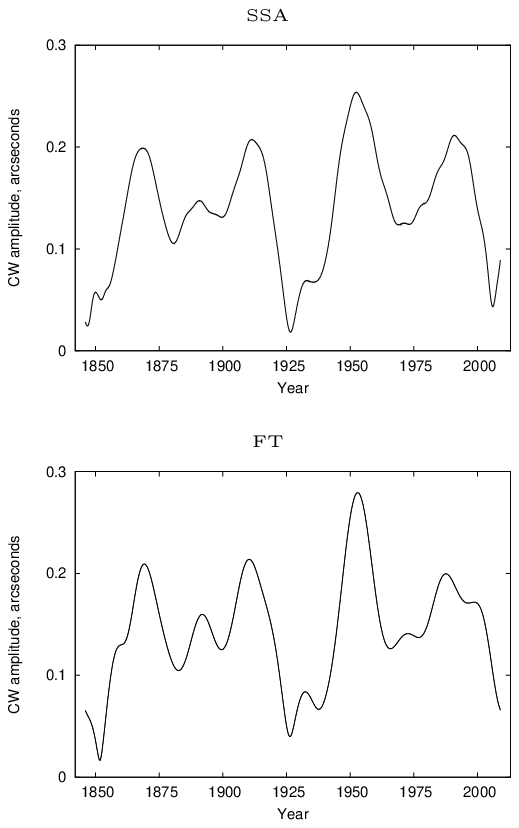}
\caption{The CW amplitude computed for  SSA-filtered and FT-filtered CW time series. Unit: mas.
One can see similar behavior of the CW amplitude obtained for both series,
with some differences near the ends of the interval.
However, three deep minima below 0.05 mas around 1850, 1925 and 2005 coincide in both cases.}
\label{fig:ampl}
\end{figure}

The CW amplitude variations thus obtained are shown in Fig.~\ref{fig:ampl} for
two CW series. We can see that both CW series show very similar behavior of the
CW amplitude, with some differences near the ends of the interval. In both CW
series, three deep minima of the amplitude below 0.05 mas around 1850, 1925 and
2005 are unambiguously detected.

Evaluation of the CW phase is a more complicated task.
Two methods of investigation of the CW phase variations were examined.
The first method was developed by Malkin (2007) for Free Core Nutation modelling.
The computations are made in two steps.
In the first one the wavelet analysis is applied to both CW series to get the period variations.
To perform the wavelet transform (WT) we used the program WWZ developed by the American
Association of Variable Star Observers\footnote{http://www.aavso.org/}.
Theoretical background of this method can be found in Foster (1996).
Since, as discussed above, we cannot separate by a mathematical tool the phase variations
from the period variations, we consider the WT output as apparent period variations $P(t)$.
Then we can compute the phase variations $\Phi(t)$ as
\begin{equation}
\Phi(t) = \int\limits_{t_0}^t {\displaystyle\frac{2\pi}{P(t)}\,dt} + \Phi_0 \,,
\label{eq:phase2}
\end{equation}
here $\Phi_0$ is the parameter to be adjusted.

The second method we used to evaluate the phase variations is the Hilbert transform (HT).
For this work we used the function {\tt hilbert} from the MATLAB Signal Processing Toolbox.

\begin{figure}[!ht]
\centering
\epsfclipon \epsfxsize=\hsize \epsffile{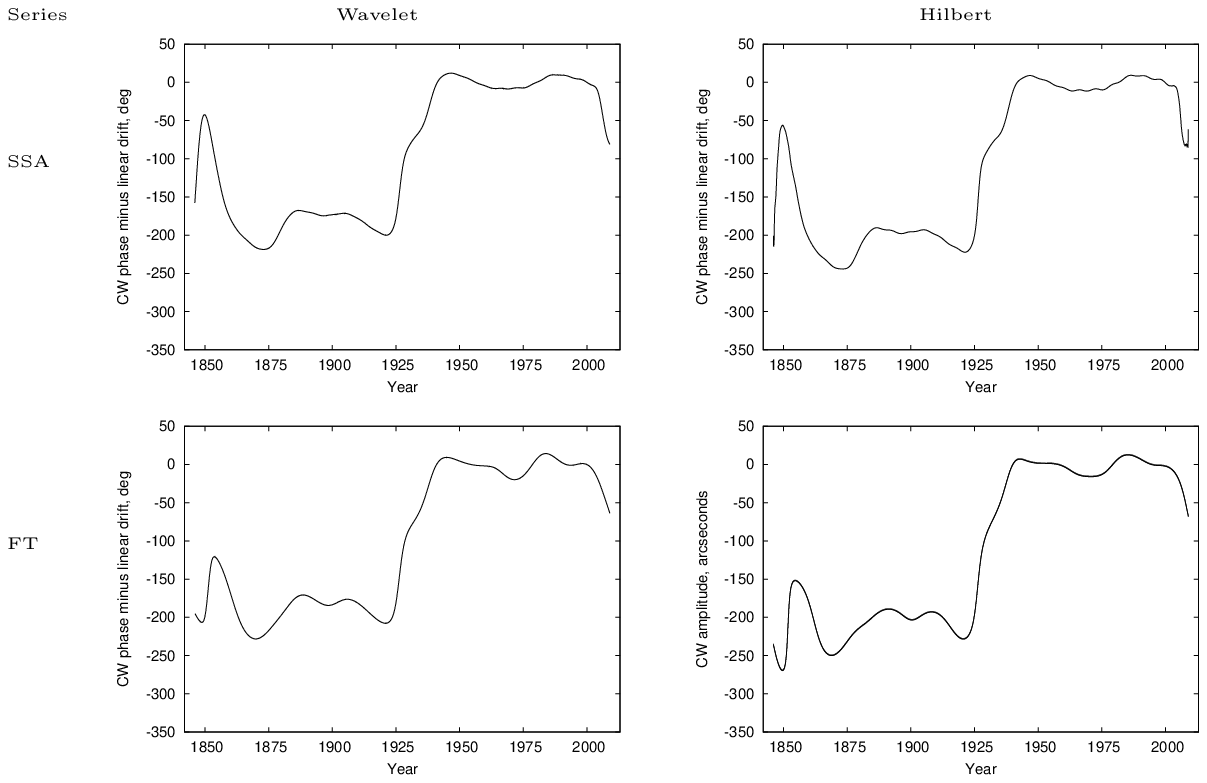}
\caption{The CW phase variations computed for SSA-filtered and FT-filtered CW series
using WT and GT (see section 3 for details). Unit: degrees.
One can see similar behavior of the CW phase obtained in all four variants,
with some differences near the ends of the interval.
Three large phase jumps in 1850s, 1920s and 2000s are clearly visible in all the cases.}
\label{fig:phi}
\end{figure}

Thus we computed the CW phase variations for two methods of the PM series
filtering and two methods of the CW phase evaluation. The results of the
computation of the CW phase variations after removing the linear trend
corresponding to $P_0$ are shown in Fig.~\ref{fig:phi}. One can see similar
behavior of the CW phase obtained in all four variants, with some differences
near the ends of the interval. However, substantial phase jumps in the 1850s
and 2000s are clearly visible in all the cases, and their epochs are
contemporary with the minima of the CW amplitude as shown in
Fig.~\ref{fig:ampl}. So, we can conclude that the well-known event in the 1920s
of the simultaneous deep minimum of the CW amplitude and the large phase jump
may be not unique.

%%%%%%%%%%%%%%%%%%%%%%%%%%%%%%%%%%%%%%%%%%%%%%%%%%%%%%%%%%%%%%%%%%%%%%%%%%%%%%

\section{Conclusion}
\label{sect:concl}

In this paper, we have investigated the whole 163-year PM series available
from the IERS with the main goal to reveal and evaluate the major CW phase jumps.
To improve the reliability of the results we used several analysis
methods. To extract the CW signal from this series SSA-based and FT-based
digital filters were applied. The two CW series thus obtained were used to
investigate the CW amplitude and phase variations. While computation of the CW
amplitude is straightforward, computation of the CW phase is not such an
unambiguous procedure. Two methods, WT and HT, have been applied to evaluate the CW
phase variations.

All the methods used gave very similar results, with some differences at the
ends of the interval. These discrepancies can be explained by different edge
effects of the methods used, but they can hardly discredit the final conclusion
that can be made from this study about existence of two epochs of deep CW
amplitude decrease around 1850 and 2005, which are also accompanied by a large
phase jump, like the well-known event in the 1920s. Thus, the latter seems to
be not unique anymore.

Unfortunately, both periods of the phase disturbances found in this paper are
located at the edges of the interval covered by the IERS EOP series.
As for the end of the interval, the next decade will allow us to quantify the phase jump
in the beginning of the 21st century more accurately.
On the other hand, a supplement study
seems to be extremely important to try improve our knowledge of the PM in the
19th century, including an extension of the PM series in the past. As
investigated in detail by Sekiguchi (1975), there are several latitude series
obtained in the first half of 19th century, which can be used to extend the
IERS C01 PM series back to the 1830s. However, most of the observations are of
rather poor quality to be used directly in the computation of an extended PM
series. Clearly, this material is worth revisiting and reprocessing using
HIPPARCOS and later GAIA star positions and proper motions. An attentive and
critical look into processing of the historical observations also could improve
our knowledge of the latitude variations in 19th, and maybe even 18th, century.

%%%%%%%%%%%%%%%%%%%%%%%%%%%%%%%%%%%%%%%%%%%%%%%%%%%%%%%%%%%%%%%%%%%%%%%%%%%%%%

{\bf Acknowledgments.}
This research has made use of the IERS PM series available at \verb"http://www.iers.org".

%%%%%%%%%%%%%%%%%%%%%%%%%%%%%%%%%%%%%%%%%%%%%%%%%%%%%%%%%%%%%%%%%%%%%%%%%%%%%%

\bigskip
\bigskip
\noindent{\Large\bf References}
\bigskip

\leftskip=\parindent
\parindent=-\leftskip

Chandler, S. C., On the variation of the latitude, I, {\it Astron. J.}, {\bf 11}, 59--61, 1891a.

Chandler, S. C., On the variation of the latitude, II, {\it Astron. J.}, {\bf 11}, 65--70, 1891b.

Fedorov, E. P. and , Ya. S. Yatskiv, The cause of the apparent ``bifurcation'' of the free nutation period,
 {\it Soviet Astron.}, {\bf 8}, 608--611, 1965

Foster, G., Wavelets for period analysis of unevenly sampled time series, {\it Astron. J.}, {\bf 112}, 1709--1729, 1996.

Gibert, D. and J.-L. Le Mou\"el, Inversion of polar motion data: Chandler wobble, phase jumps, and geomagnetic jerks,
 {\it J.~Geophys. Res.}, {\bf 113}, B10405, 2008.

Guinot, B., Astron. Astrophys., The Chandlerian Wobble from 1900 to 1970, {\bf 19}, 207--214, 1972.

Guo, J. Y., H. Greiner-Mai, L. Ballani, H. Jochmann, and C. K. Shum, On the double-peak spectrum of the Chandler wobble,
 {\it J.~Geodesy}, {\bf 78}, 654--659, 2005

Lambeck, K., The Earth's Variable Rotation: Geophysical causes and consequences. {\it Cambridge Univ. Press}, 1980.

Malkin, Z. M., Empiric Models of the Earth's Free Core Nutation, {\it Solar System Research}, {\bf 41}, 492--497, 2007.

Miller, N., {\it Izvestiya Vysshikh Uchebnykh Zavedeniy: Geodeziya i Aerofotos'emka}, On variations of the amplitude and phase of the Chandler wobble,
 No.~5, 48, 2008. (in Russian)

Munk, W. H. and G. J. F. MacDonald, The Rotation of the Earth. {\it Cambridge Univ. Press}, 1960.

Orlov, A., {\it Doklady AN SSSR}, {\bf 43}, On triaxiality of the Earth, 343--345, 1944. (in Russian)

Sekiguchi, N., On the latitude variations of the interval between 1830 and 1860, {\it J. Geod. Soc. Japan}, {\bf 21}, 131--141, 1975.

Sekiguchi, N., An interpretation of the multiple-peak spectra of the polar wobble of the Earth, {\it Publ. Astron. Soc. Japan}, {\bf 28}, 277, 1976

Shirai, T, T. Fukushima, and Z. Malkin, Detection of phase disturbances of free core nutation of the Earth and their concurrence with geomagnetic jerks,
 {\it Earth Planets Space}, {\bf 57}, 151, 2005.

Schuh, H., S. Nagel, and T. Seitz, Linear drift and periodic variations observed in long time series in polar motion,
 {\it J. Geodesy}, {\bf 74}, 701, 2001.

Vondr\'ak, J., Is Chandler frequency constant, in {\it The Earth's Rotation and Reference Frames for Geodesy and Geodynamics,
 Proc. IAU Symp. 128}, ed. A.~K.~Babcock \& G.~A.~Willis, 359, 1988.

Vorotkov, M. V., V. L. Gorshkov, N. O. Miller,  and E. Ya. Prudnikova, The investigation of the main components in the polar motion of the Earth,
 {\it Izvestiya GAO v Pulkove}, No.~216, 406--414, 2002. (in Russian)

\end{document}